\documentclass{sig-alternate-05-2015}

\usepackage{booktabs}
\usepackage{mathtools}
\usepackage{multirow}
\usepackage{url}
\usepackage{xspace}
\usepackage{bm}
\usepackage{graphicx}
\usepackage[center]{subfigure}
\usepackage{ctable}
\usepackage{tabularx}
\usepackage{stfloats}

\newcounter{swacounter}
\DeclareRobustCommand{\swa}[1]{\textbf{/* #1 (swa) */}\stepcounter{swacounter}\typeout{LaTeX Warning: swa comment \theswacounter: #1 (line \the\inputlineno)}}

\newcounter{dhecounter}
\DeclareRobustCommand{\dhe}[1]{\textbf{/* #1 (dhe) */}\stepcounter{dhecounter}\typeout{LaTeX Warning: dhe comment \thedhecounter: #1 (line \the\inputlineno)}}

\newcounter{mstcounter}
\DeclareRobustCommand{\mst}[1]{\textbf{/* #1 (mst) */}\stepcounter{mstcounter}\typeout{LaTeX Warning: mst comment \themstcounter: #1 (line \the\inputlineno)}}

\newcounter{lescounter}
\DeclareRobustCommand{\les}[1]{\textbf{/* #1 (les) */}\stepcounter{lescounter}\typeout{LaTeX Warning: les comment \thelescounter: #1 (line \the\inputlineno)}}

\iftrue
    \renewcommand{\swa}[1]{}
\fi

\begin{document}
\setcopyright{acmcopyright}

\conferenceinfo{Under review for WWW'17}{April 3--7, 2017, Perth, AU}

\title{How Users Explore Ontologies on the Web: \\A Study of NCBO's BioPortal Usage Logs}

\numberofauthors{5} %

\author{
\alignauthor
Simon Walk\\
       \affaddr{Stanford University \& Graz University of Technology}\\
       \email{walk@stanford.edu}
\alignauthor
Lisette Esp\'in-Noboa\\
       \affaddr{GESIS \& University of Koblenz-Landau}\\
       \email{lisette.espin@gesis.org}
\alignauthor
Denis Helic\\
       \affaddr{Graz University of Technology}\\
       \email{dhelic@tugraz.at}
\and  %
\alignauthor Markus Strohmaier\\
       \affaddr{GESIS \& University of Koblenz-Landau}\\
       \email{markus.strohmaier@gesis.org}%
\alignauthor Mark A. Musen\\
       \affaddr{Stanford University}\\
       \email{musen@stanford.edu}
}

\maketitle
\begin{abstract}

Ontologies in the biomedical domain are numerous, highly specialized and very expensive to develop. Thus, a crucial prerequisite for ontology adoption and reuse is effective support for exploring and finding \emph{existing} ontologies. Towards that goal, the National Center for Biomedical Ontology (NCBO) has developed BioPortal---an online repository designed to support users in exploring and finding more than $500$ existing biomedical ontologies. In $2016$, BioPortal represents one of the largest portals for exploration of semantic biomedical vocabularies and terminologies, which is used by many researchers and practitioners. 
While usage of this portal is high, we know very little about how exactly users search and explore ontologies and what kind of usage patterns or user groups exist in the first place. %
Deeper insights into user behavior on such portals can provide valuable information to devise strategies for a better support of users in exploring and finding existing ontologies, and thereby enable better ontology reuse. 
To that end, we study and group users according to their browsing behavior on BioPortal using data mining techniques. %
Additionally, we use the obtained groups to characterize and compare exploration strategies across ontologies.
In particular, we were able to identify seven distinct browsing-behavior types, which all make use of different functionality provided by BioPortal. For example, \textit{Search Explorers} make extensive use of the search functionality while \textit{Ontology Tree Explorers} mainly rely on the class hierarchy to explore ontologies.
Further, we show that specific characteristics of ontologies influence the way users explore and interact with the website. Our results may guide the development of more user-oriented systems for ontology exploration on the Web. %
\end{abstract}
\vspace{-4pt}





\section{Motivation}

Facilitating reuse on the Semantic Web requires support for effectively finding and exploring existing semantic resources such as ontologies or taxonomies. Particularly in the biomedical domain, where ontologies are highly specialized, costly and often created collaboratively by large groups of domain experts, \textit{identifying and finding existing semantic vocabularies and terminologies} is crucial. To support researchers and practitioners in this task, the National Center for Biomedical Ontology (NCBO) has developed BioPortal\footnote{\url{http://bioportal.bioontology.org}}---\emph{the world's most comprehensive online repository of biomedical ontologies} \cite{salvadores2013bioportal, musen2012national,whetzel2011bioportal, noy2009bioportal}.

\begin{table*}[!ht]
\center
\scriptsize
\caption{Overview of all actions available in the BioPortal user interface.}
\begin{tabular}{ p{2.2cm} | p{8.5cm} | p{5.5cm} }
\toprule
\multicolumn{1}{c}{} & \multicolumn{1}{c}{\textbf{Action Labels}} & \multicolumn{1}{c}{\textbf{Description}}  \\\midrule
\multirow{3}{2.2cm}{\centering \textbf{Main Page}} & \textit{Browse Main Page}, \textit{Browse Ontologies}, \textit{Browse Search}, \textit{Browse Help}, \textit{Browse Mappings}, \textit{Browse Recommender}, \textit{Browse Annotator}, \textit{Browse Resource Index}, \textit{Browse Projects}, \textit{Browse Notes} & \multirow{3}{5.5cm}{\centering Browsing main areas and functionality of BioPortal.} \\\midrule
\multirow{6}{2.2cm}{\centering \textbf{Ontology Page}} & \textit{Ontology Summary}, \textit{Browse Ontology Classes}, \textit{Browse Ontology Class}, \textit{Browse Ontology Class Tree}, \textit{Browse Ontology Mappings}, \textit{Ontology Analytics}, \textit{Browse Ontology Widgets}, \textit{Browse Ontology Visualization}, \textit{Browse Ontology Notes}, \textit{Browse Ontology Properties}, \textit{Browse Widgets}, \textit{Browse Ontology Property Tree}, \textit{Browse Class Notes} & \multirow{6}{5.5cm}{\centering Labels for actions that can be performed while browsing a specific ontology on BioPortal.} \\\midrule
\multirow{2}{2.2cm}{\centering \textbf{Edit Content}} & \textit{Create Ontology Submission}, \textit{Validate Ontology File}, \textit{Virtual Appliance Download}, \textit{Browse Ontology Submission} & \multirow{2}{5.5cm}{\centering Actions triggered when uploading new versions of ontologies.}\\\midrule
\multirow{1}{2.2cm}{\centering \textbf{User Account}} & \textit{Login}, \textit{Log-Out}, \textit{Sign-Up}, \textit{Lost Password}, \textit{Browse Account}, \textit{Feedback} & \multirow{1}{5.5cm}{\centering Actions to manage accounts on BioPortal.} \\\midrule
\multirow{1}{2.2cm}{\centering \textbf{Control Term}} & \textit{BREAK} & \multirow{1}{5.5cm}{\centering $\geq$30 minutes inactivity between two actions.} \\
\bottomrule
\end{tabular}
\label{tab:click types}
\vspace{-5pt}
\end{table*}

\noindent\textbf{Problem.} In this paper, we want to shed light on how users are exploring ontologies and terminologies in BioPortal. Without a deep understanding of how to facilitate the exploration of existing ontological resources on the Web, increasing the reuse of ontologies on the Semantic Web will remain an elusive goal. Towards that goal, we want to (i) cluster traces of user interactions to obtain different exploration behavior types and (ii) investigate ontologies in terms of their user behavior. New insights into user behavior on such portals can potentially provide actionable information to devise strategies that better support researchers and practitioners in exploring existing ontologies, and thereby may enable better reuse of ontological content in general. 

\noindent\textbf{Approach.} To identify different types of user behavior on BioPortal, we adopt a clustering approach based on (i) dynamic browsing features extracted from the Apache logs of BioPortal and on (ii) calculating stationary distributions from first order Markov chain representations of user transitions. We demonstrate the effectiveness of our approach and investigate if and to what extent exploration strategies of users differ when exploring biomedical ontologies on BioPortal. In addition, we use Principal Component Analysis (PCA) to visually inspect the obtained clusters.

\textbf{Contributions.} We present an approach for identifying different user behavior types in ontology exploration log data using stationary distributions from first order Markov chains. We provide evidence for a total of $7$ distinct browsing-behavior types on BioPortal and offer an interpretation of their relevance and meaning. We use these different types of exploration behavior to characterize ontologies in terms of the different user interactions they attract. We find that particular characteristics of semantic resources, as represented on BioPortal, influence how users interact with them. For example, BioPortal lacks the ability to visualize flat ontologies (i.e., ontologies without a structured hierarchy), which fuels the emergence of diverse exploration strategies as identified by our approach. Overall, our work advances our understanding of ontology exploration behavior on prominent repositories of biomedical ontologies on the Web.

\section{Related Work}

\noindent\textbf{Biomedical Ontology Repositories.}
There exists a vast variety of repositories for ontologies and taxonomies with similarly diverse core interests. We provide a brief overview of the most important projects from the biomedical domain and describe how they are related to BioPortal.
The Open Biomedical Ontologies (OBO) Foundry \cite{smith2007obo} initiative collects and maintains a set of different ontologies, which are specifically designed for interoperability. The whole library of ontologies is available on GitHub\footnote{\url{http://obofoundry.org}} and developers of ontologies aspire to have their content included in the OBO Foundry. All ontologies of the OBO Library are automatically imported into BioPortal.

The European Bioinformatics Institute \cite{jupp2015new} maintains the Ontology Lookup Service which provides an alternative me\-cha\-nism for accessing content in the OBO Library.

The Unified Medical Language System (UMLS) \cite{bodenreider2004unified} is a collection of biomedical terminologies and ontologies distributed by the National Library of Medicine.  Terms in UMLS are mapped to one another through the UMLS Meta\-the\-saurus. Similarly to OBO Foundry, the content of UMLS is also available for exploration in BioPortal.

The majority of research about (biomedical) ontology re\-po\-sitories focuses on the selection and maintenance (including provenance) of the included ontologies and taxonomies. In this paper, we present analyses that complement existing research by studying aspects about the interaction behavior of users on BioPortal~\cite{salvadores2013bioportal, musen2012national,whetzel2011bioportal, noy2009bioportal}---one of the largest biomedical ontology and taxonomy repositories on the Web. 

\begin{figure*}[!t]
\centering
\subfigure[Seconds between requests]{\label{fig:bp_timediff:seconds}\includegraphics[width=0.24\textwidth]{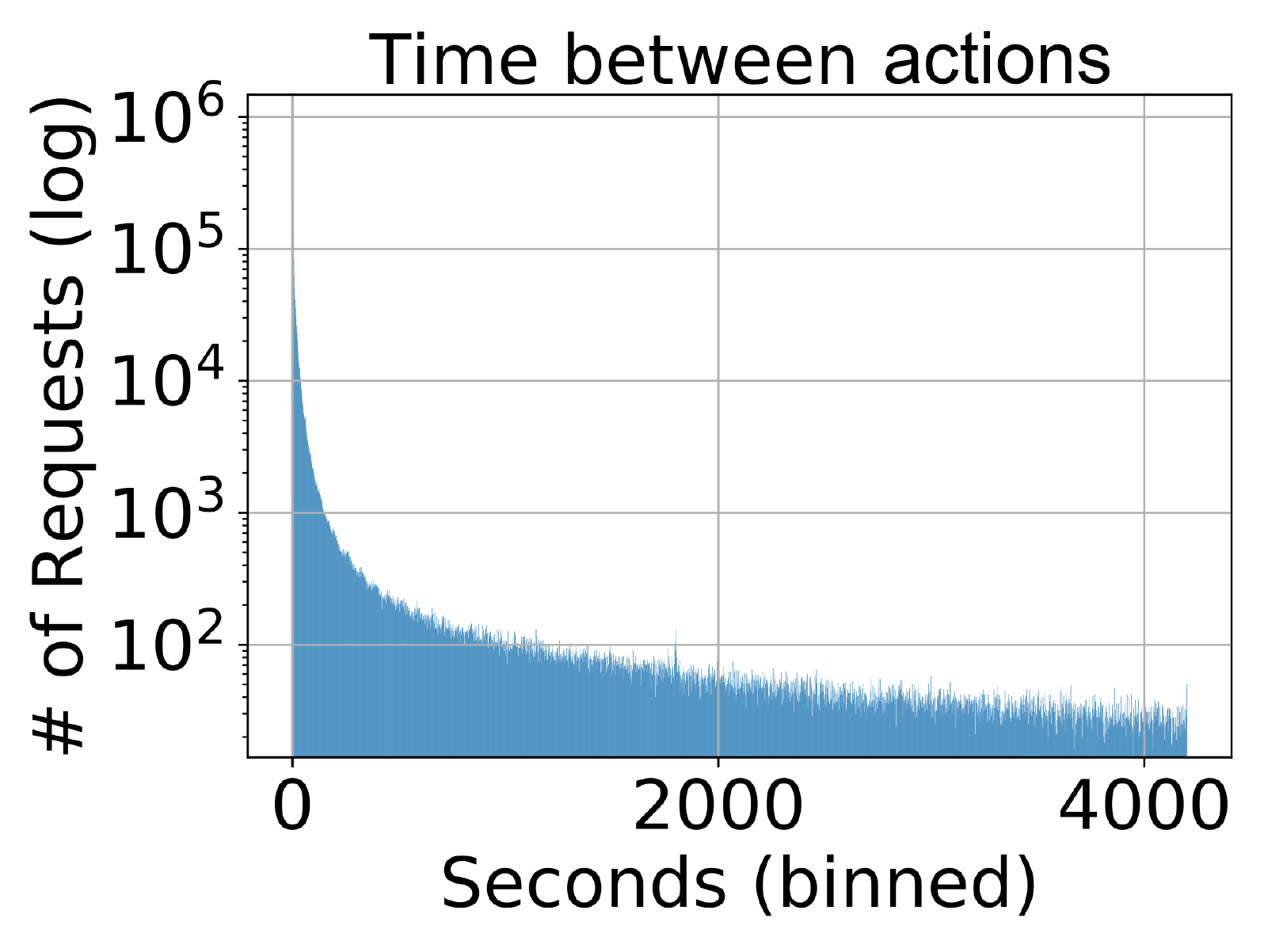}}
\subfigure[Requests per IP]{\label{fig:bp_timediff:cr_per_ip}\includegraphics[width=0.24\textwidth]{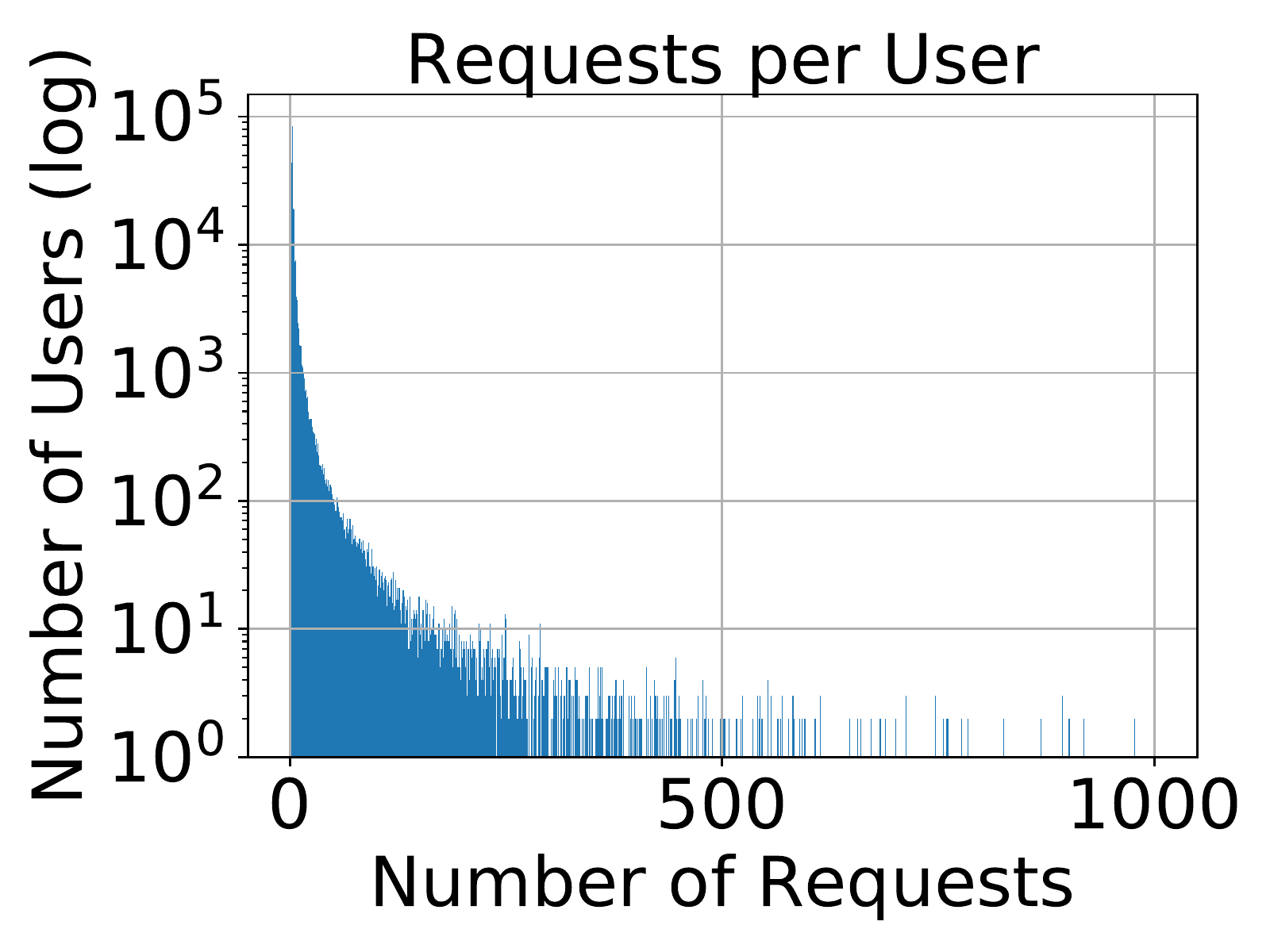}}
\subfigure[Requested ontologies per user]{\label{fig:bp_timediff:ont_per_user}\includegraphics[width=0.24\textwidth]{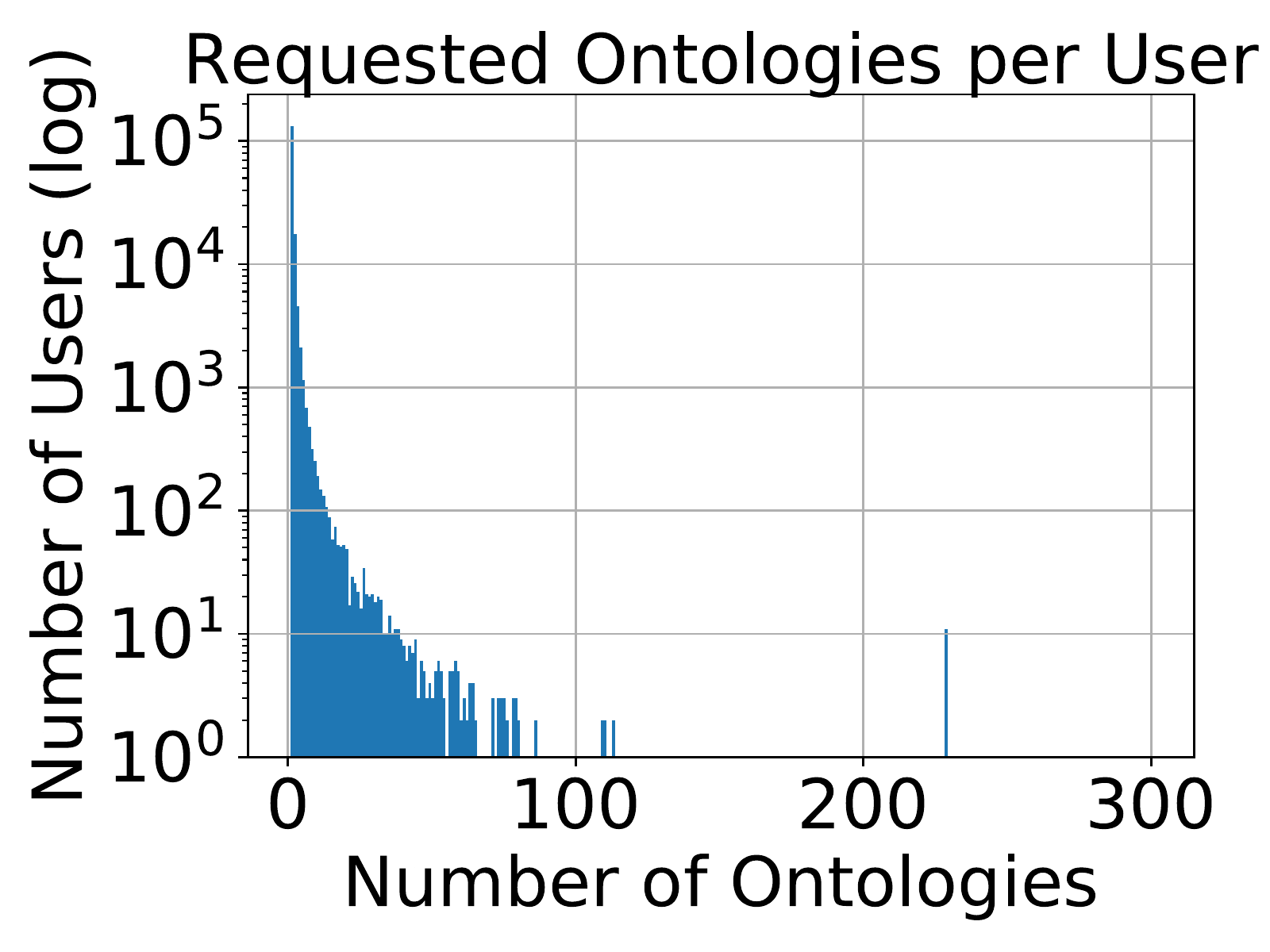}}
\subfigure[Requests per session]{\label{fig:bp_timediff:cr_per_session}\includegraphics[width=0.24\textwidth]{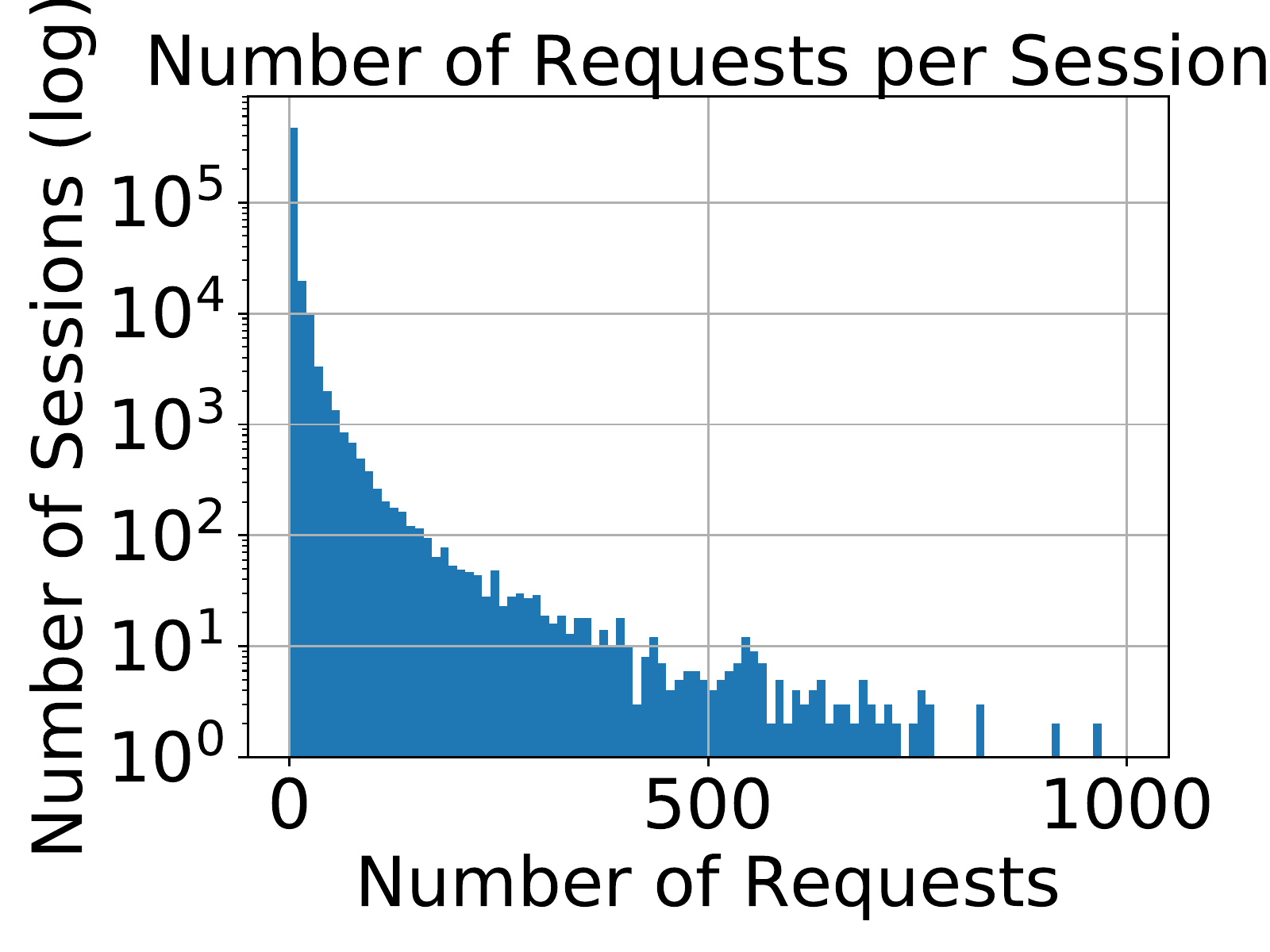}}
\caption{
\textbf{\textit{BioPortal Characteristics}.} Entities on the $x$-axes for all figures are binned according to the displayed values and cut-off at their maximum values for reasons of readability. The seconds between requests (a) are heavy-tailed, meaning that the majority of requests are conducted within short periods of times. %
The number of request actions per IP (b), a proxy for users on BioPortal, is heavy-tailed. 
The majority of users have visited only a total of one or two ontologies and (c) the majority of sessions exhibit a small number of requests (d).
}
\label{fig:bp_timediff}
\vspace{-10pt}
\end{figure*}

\noindent\textbf{User Interactions on the (Semantic) Web.}
A lot of research has addressed the tasks, actions, tools and processes required for developing and evaluating ontologies~\cite{gomez2004ontological}. This has led the Semantic Web community to develop a variety of different guidelines and methodologies \cite{simperl2014collaborative} %
or compile best practices \cite{noy2001ontology} for engineering ontologies. %

Engineering ontologies collaboratively can potentially lead to structural conflicts since several users can edit ontologies simultaneously. To circumvent this, Hellman and Gal~\cite{Hellman2005} propose locking mechanisms based on a graph depiction of the concept dependencies. Pesquita et al. \cite{10.1371/journal.pcbi.1002630} leveraged the location and specific structural features to show that these can be used to determine if and where the next change is going to take place in the Gene Ontology\footnote{\url{http://www.geneontology.org}}. 

To learn more about the impact and quality of collaboration in collaborative ontology-engineering projects, Stroh\-maier et al. \cite{j.websem333} investigated the hidden social dynamics that take place in such projects from the biomedical domain and provided new metrics to quantify various aspects of the collaborative engineering processes. User roles have been also studied from many different angles. For example, Falconer et al.~\cite{conf/kcap/FalconerTN11} investigated and classified users according to different roles in collaborative ontology-en\-gi\-neering projects. In 2014, Van Laere et al.~\cite{van2014method} used $K$-means and the GOSPL methodology to classify users by clustering interactions that users engage in, while engineering an ontology. Wang et al. \cite{wang2013analysis} used association-rule mining on the change-logs of collaborative ontology-engineering projects, and showcased the utility of the identified editing patterns in a prediction experiment. Walk et al.~\cite{walk-iswc} studied user editing trails of ontology-engineering projects by defining and comparing a set of hypotheses about how users edit ontologies. 

Further, there has been increasing interest on analyzing and categorizing human sequences in the past, including mobility~\cite{espin2016discovering}, page access~\cite{fortuna2011user, kumar2010characterization}, edit ~\cite{walk-iswc,Walk201551,DBLP:journals/jbi/WalkSSTMN14} as well as click stream logs \cite{Singer:2015:HBA:2736277.2741080}. %
Additionally, Markov models of various orders have been employed~\cite{borges, lempel, pirolli, deshpande, sen, DBLP:conf/cikm/WalkSS14} to model and predict clicks or interactions of users on the (Semantic) Web. 
Recently, Chierichetti et al.~\cite{chierichetti} conducted an analysis that questioned if a first-order model best represents the navigation behavior of humans on the Web. In this direction, the work in Lakshimarayan et al.~\cite{lakshminarayan2016modeling} demonstrates that higher order Markov chains can be used to model Web browsing behavior and predict user intent towards buying specific products. 

Making sense and interpreting such sequential data is not a trivial task. The work in Hoxha et al.~\cite{hoxha2012enabling} proposes a platform to analyze user browsing patterns by semantically formalizing the usage logs of Web data, providing means to query and retrieve sessions of users that satisfy certain semantic and temporal conditions. 
Research has also been focused on human behavior beyond browsing patterns and study reading behavior of users in Wikipedia as well as the stability of the reading patterns per user~\cite{lehmann2014reader}. A more generic approach is proposed in~\cite{Singer:2015:HBA:2736277.2741080} where the authors present HypTrails---a Bayesian methodology that allows researchers to compare and rank hypotheses about digital trails on the Web, which has also been applied on the edit logs of collaborative ontology-engineering projects by Walk et al.~\cite{walk-iswc}.

To the best of our knowledge, analyzing how users explore ontologies in the context of ontology repositories on the Web has not been addressed as in-depth before. %

\section{Materials \& Methods}

\subsection{BioPortal}
\label{subsec:bioportal}
BioPortal---an online biomedical ontology and taxonomy repository---was created by the National Center for Biomedical Ontology (NCBO). The main goals of NCBO involve not only the creation and maintenance of a comprehensive repository of biomedical ontologies and terminologies, but also the task of building novel tools and Web services to enable and augment the (re)use of all stored semantic vocabularies and terminologies in clinical and translational research. 
BioPortal currently contains more than $500$ biomedical ontologies, and supports a wide range of Web services, such as using ontologies for annotating resources and generating value sets or specific ontology views. Additionally, BioPortal allows users to explore ontological content not only by using a standard tree browser, but it can also visualize resources using custom tailored widgets, which help users in comprehending the complexity of large biomedical resources. Similarly, the website provides functionality to explore mappings between ontologies, which can be used to directly compare the use of related terms as well as the overlap between different ontologies.

\begin{table}[!b]
\vspace{-10pt}
\center
\small
\caption{Characteristics of the BioPortal dataset.}
\begin{tabular}{ l | c }\toprule
\multicolumn{1}{c|}{\textbf{Feature}} & \textbf{Value}  \\\midrule
Unique IPs & $215,908$ \\
Unique IPs (with $\geq2$ requests) & $168,008$ \\
Ontologies & $1,818$ ($517$) \\
Actions (ca.) & $2.52$M \\\midrule
Sessions (time between requests $\geq30$ mins) & $513,659$ \\
$1$-request sessions & $165,543$ \\
\# requests in $\geq2$-request sessions & $2,36$M \\
Average/Median session duration & $217$s/$0$s \\\midrule
First request & 2015/12/31 \\
Last request & 2016/07/19 \\
Observation period (ca.) & 7 months \\\bottomrule
\end{tabular}
\label{tab:dataset details}
\end{table}

\begin{table*}[!ht]
\center
\scriptsize
\caption{\textit{Example of a session.} Using the Apache logs of BioPortal, we can chronologically order the requests of each user (identified by IP) individually, assign action labels to each request and create a sequence of actions, which can not only be used to fit first-order Markov chains, but also to calculate stationary distributions for each user.}
\begin{tabular}{ l | l | l | c }\toprule
\multicolumn{1}{c|}{\textbf{Timestamp}} & \textbf{Type} & \multicolumn{1}{c|}{\textbf{Request}} & \multicolumn{1}{c}{\textbf{Action Labels (Sequence Step)}} \\\midrule
\textit{2016-03-14 09:07:32} & GET & / & Browse Main Page (1)\\
\textit{2016-03-14 09:07:46} & GET & /login?redirect=http\%3A\%2F\%2Fbioportal.bioontology.org\%2F & Login  (2)\\
\textit{2016-03-14 09:07:48} & POST & /login & Login (3)\\
\textit{2016-03-14 09:07:50} & GET & / & Browse Main Page (4)\\
\textit{2016-03-14 09:08:04} & GET & /ontologies/MCCV & Ontology Summary (5)\\
\textit{2016-03-14 09:08:22} & GET & /ontologies/MCCV/submissions/new & Create Ontology Submission (6)\\
\textit{2016-03-14 09:09:34} & POST & /ontologies/MCCV/submissions & Create Ontology Submission (7)\\
\textit{2016-03-14 09:09:59} & GET & /ontologies/success/MCCV & Create Ontology Submission (8)\\
\textit{2016-03-14 09:10:14} & GET & /ontologies/MCCV & Ontology Summary (9)\\
\midrule
\multicolumn{4}{p{17.2cm}}{\tiny{\textbf{Sequence:} Browse Main Page --> Login --> Login --> Browse Main Page --> Ontology Summary --> Create Ontology Submission --> Create Ontology Submission --> Create Ontology Submission --> Ontology Summary}}
\end{tabular}
\label{tab:click-sequence}
\vspace{-5pt}
\end{table*}

\begin{table*}[bp]
\vspace{-5pt}
\center
\footnotesize
\caption{\textit{Stationary Distribution and Page Views Illustration.} Given a vector of page views (2xA, 2xB and 2xC) the stationary distribution depends on the order of the appearance of each state in the sequence. For example, given a sequence of ``ABCABC'', we create a weighted matrix by counting the transitions between the three states A, B and C. Normalizing each row produces the transition matrix $\bm{P}$, which we use to calculate the stationary distribution $\bm{\pi}$ (top row). For a user with the same page views, but a different sequence, we obtain a different stationary distribution (bottom row). The static page views remain unaffected by the ordering of the sequence, as only the total number of occurrences of each state is considered.}
\begin{tabular}{ c  c  c  c c }\toprule
\textbf{Sequence} & \textbf{Weighted Matrix} & \textbf{Transition Matrix} & \textbf{Stationary Distribution} & \textbf{Page Views}\\\midrule
ABCABC &

$\bm{A} = \begin{pmatrix}  
  0 & 2 & 0 \\
  0 & 0 & 2 \\
  1 & 0 & 0 
 \end{pmatrix} $ & $\bm{P} = \begin{pmatrix} 
  0 & 1 & 0 \\
  0 & 0 & 1 \\
  1 & 0 & 0 
\end{pmatrix} 
$ & 
$ \bm{\pi} = \begin{pmatrix} 
 0.3533  \\
 0.3356  \\
 0.3111  
 \end{pmatrix}
$ & 
$ \begin{pmatrix} 
 2  \\
 2  \\
 2  
 \end{pmatrix}$
\\
\midrule
AABBCC &

$\bm{A} = \begin{pmatrix}  
  1 & 1 & 0 \\
  0 & 1 & 1 \\
  0 & 0 & 1 
 \end{pmatrix} $ & $\bm{P} = \begin{pmatrix} 
  0.5 & 0.5 & 0 \\
  0 & 0.5 & 0.5 \\
  0 & 0 & 1 
\end{pmatrix} 
$ & 
$ \bm{\pi} = \begin{pmatrix} 
 0.5229  \\
 0.2916  \\
 0.1855  
 \end{pmatrix}
$ & 
$ \begin{pmatrix} 
 2  \\
 2  \\
 2  
 \end{pmatrix}$
\\
\bottomrule
\end{tabular}
\label{tab:sd_example}
\vspace{-10pt}
\end{table*}

\subsection{Data Acquisition \& Preprocessing}
\label{subsec:preprocessing}

For the analyses presented in this paper, we have parsed all the requests stored in the Apache logs of BioPortal from 12/31/2015 to 07/19/2016 (around 7 months; cf. Table~\ref{tab:dataset details}). 
Each line in the request logs contains information (among others) about the IP, the timestamp, the actual resource that was requested and the useragent, which is used to provide additional information about the requester, such as browser, operating system or name of the bot/spider, for each request.

\noindent\textbf{Preprocessing.} In a first step, we reduced a total of roughly $51.8$M entries ($7$ months) to $16.7$M by removing all requests that were conducted by bots, spiders and other automatic scripts via the combination of existing useragent and IP blacklists as well as through manual inspections.

Then, we reduced our dataset to only include requests, which were triggered by users and involve an actual interaction with the website. For example, if users click on an ontology name, many automatic requests (AJAX calls) are triggered, which do not represent interactions of users, and are hence filtered from our dataset. Note that we consider all interactions that trigger a request for our analyses. For example, we consider search queries entered by users on BioPortal as interactions with the website. After preprocessing, we still have a total of $2.52$M requests (see Table~\ref{tab:click types} for the different types), generated by $215,908$ unique IP addresses over the course of $7$ months. Note that the number of different ontologies that users requested is larger than the number of stored ontologies in BioPortal. Whenever users manually type in requests in the browser, typos can occur. Further, ontologies on BioPortal are also not only accessible through their abbreviations but also through their internal IDs. In both situations, the number of unique ontologies in the request logs increases, while the number of stored ontologies remains unaffected.

\noindent\textbf{Dataset Characteristics.} The distribution of time between requests (seconds) is depicted in Figure~\ref{fig:bp_timediff:seconds}. The seconds between requests are heavy-tailed, meaning that the majority of requests are performed within $0$-$10$ seconds. %

To be able to asses if and to what extent we are able to observe power users---very active users who contribute the majority of all requests---we have grouped users (in the form of IP addresses) according to the number of requests they have contributed and visualized the results in Figure~\ref{fig:bp_timediff:cr_per_ip}. While the majority only conduct a very small number of requests (between $1$-$10$), we were able to identify $30$ users who conducted more than $5,000$ interactions on BioPortal within $7$ months. Additionally, we were able to identify that the majority of users only visit one or two ontologies on the website (cf. Figure~\ref{fig:bp_timediff:ont_per_user}).

Finally, we aggregated all requests of users into sessions. Each session contains all chronologically sorted interactions of one user, unless two interactions are apart more than $30$ minutes. In that case, the current session is closed, a new session is opened and the process is repeated. We have selected $30$ minutes as threshold, as the requests per IP are heavy-tailed and the majority of timespans between requests are within that threshold, introducing only a small number of sessions.
We were able to group all of the $2.52$M requests into a total of $513,659$ sessions, where $165,543$ only consist of a single interaction. As depicted in Figure~\ref{fig:bp_timediff:cr_per_session}, the majority of sessions in BioPortal exhibit a low number of requests, while only very few sessions last for more than $100$ requests.

\noindent \textbf{Action Label Mapping.} Finally, we created a set of regular expressions to map each request to easier-readable labels (cf. Table~\ref{tab:click types} for all the different types of labels). In general, the generated action labels can be grouped into $5$ different categories. The two most important groups of action labels are the ones accessible from the main page of BioPortal, such as \textit{Browse Ontologies} or \textit{Browse Search}, and the ones that describe actions conducted on single ontology pages, such as \textit{Browse Ontology Class} or \textit{Browse Ontology Property Tree}. 

\noindent \textbf{Action Sequence Generation.} Using the action labels we are able to extract easily readable chronological sequences of actions for each user on BioPortal (see Table~\ref{tab:click-sequence}). Whenever an IP address exhibits two or more sessions, we connect the sessions with a \textit{BREAK} state to be able to see where users stop and resume browsing BioPortal.

\subsection{Modeling Browsing Behavior}
\label{subsec:markovchains}

We model user browsing behavior on BioPortal with memoryless Markov chains. A Markov chain consists of $n$ states $s_i$ from a finite state-space $S$ with $n = |S|$ and a transition matrix $\bm{P} \in \mathbb{R}^{n \times n}$ where $p_{ij}$ defines the probability to traverse from state $s_i$ to state $s_j$. Since each row in $\bm{P}$ defines a probability distribution for each $i$ we have $\sum_j{p_{ij}}=1$. 

We represent each action on BioPortal as a single state of a Markov chain (see Table~\ref{tab:click types}). Then, an element $p_{ij}$ from the transition matrix $\bm{P}$ represents the probability of performing action $j$ after action $i$ has been performed.

\begin{figure*}[!ht]
\centering
\subfigure[Selection of $K$]{\label{fig:bp_kval}\includegraphics[width=0.49\textwidth]{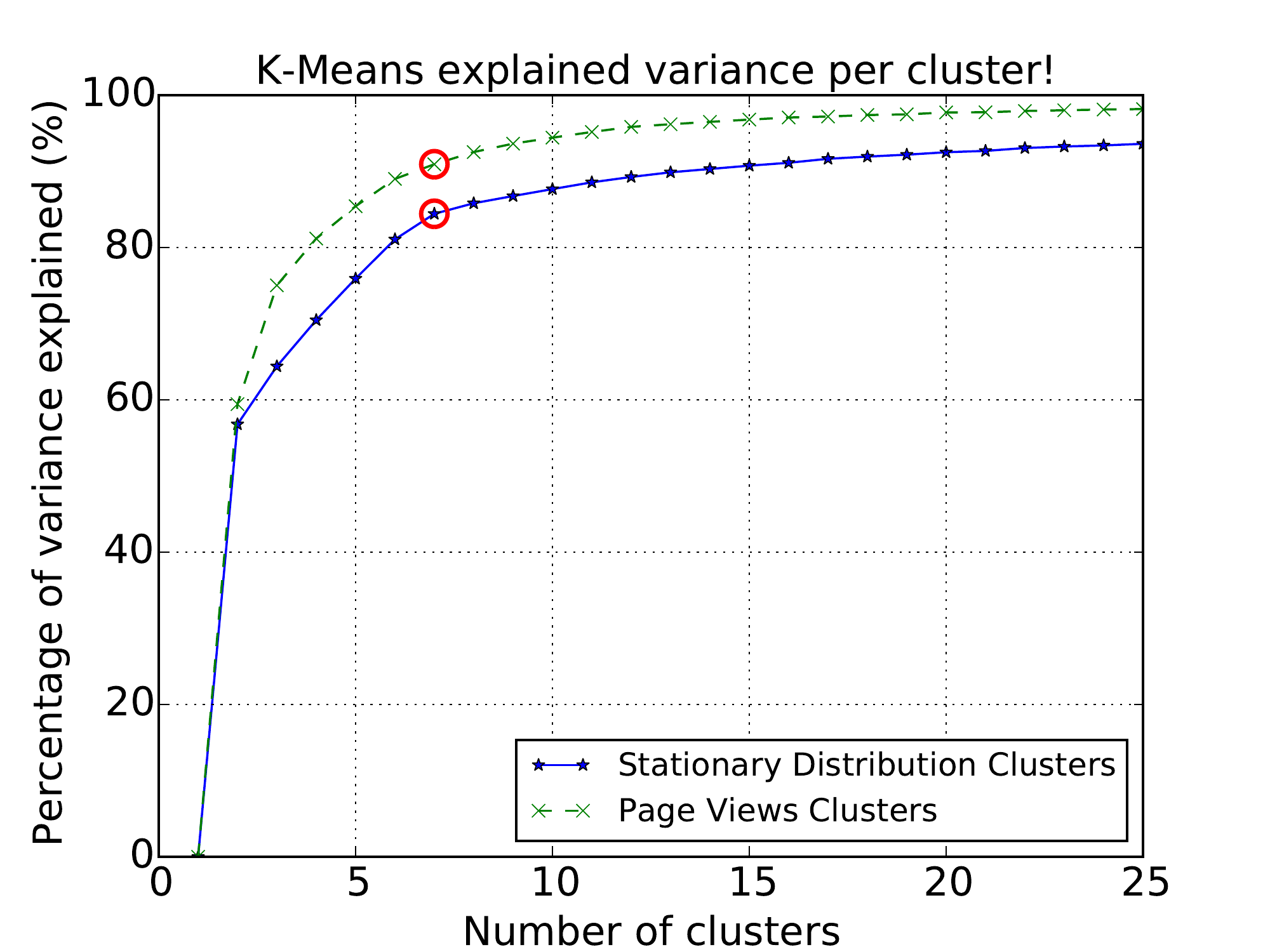}}
\subfigure[Stationary Distribution Clusters]{\label{fig:bp_statdist_clusters}\includegraphics[width=0.49\textwidth]{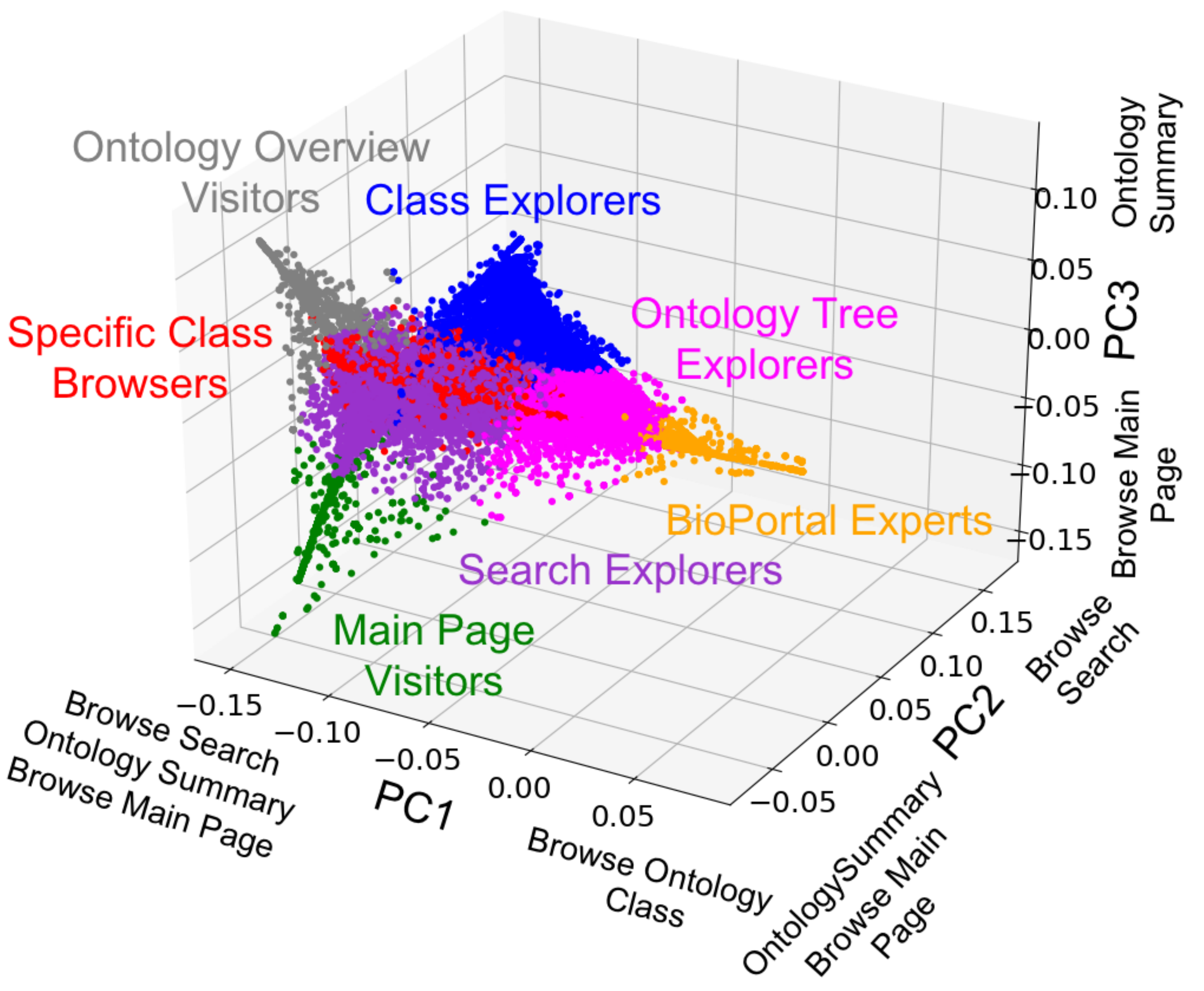}}
\caption{
\textbf{\textit{User Behavior Clusters}.} To estimate the number of clusters to investigate, we have plotted the percentage of the explained variance ($y$-axis) per cluster ($x$-axis) in (a) for the stationary distribution clusters (solid blue) and the page view clusters (dashed green). We select the number of clusters ($7$) to a value where the introduction of new clusters only minimally increases the explained variance (red circles). To further investigate the results of $K$-means, we have reduced and visualized the stationary distribution vectors of each user to three dimensions using principal component analysis in (b). Each point represents one user of our dataset, while the colors represent the corresponding clusters obtained by $K$-means (before dimensionality reduction). We provide labels for the clusters and the extremes of the axes. The latter represent the actions with the smallest and highest coefficients of the corresponding principal components (see Table~\ref{tab:pca coeffs}). 
}
\label{fig:bp_uclusts}
\vspace{-5pt}
\end{figure*}

To compare users with each other we compute $\bm{P}$ and its stationary distribution for each individual user. In our case, the stationary distribution is a probability distribution over actions, which defines how likely we will find a user performing a given action in the limit of large number of steps. Generally, the stationary distribution captures the way how users move between actions, i.e. it encodes the sequential interactions within an $n$-dimensional vector, and therefore provides additional information over simple view counts (cf. Table~\ref{tab:sd_example}).

To compute stationary distributions, we first define the weighted matrix $\bm{A} \in \mathbb{R}^{n\times n}$,
where each element $a_{ij}$ is set to the number of observed transitions between elements $i$ and $j$ in our empirical dataset. %
A transition matrix $\bm{P}$ has to satisfy certain conditions to have a stationary distribution, most notably irreducibility (each action has to be reachable from all other actions) and aperiodicity (the return times to actions have to be aperiodic).

To guarantee both irreducibility and aperiodicity, we add a teleportation factor $\alpha$ to $\bm{A}$, which (technically) would allow users to teleport between all actions with a very small probability, similar to PageRank. The teleportation factor connects each state to all others (satisfying irreducibility) including a self-loop (satisfying aperiodicity):

\vspace{-5pt}
\begin{equation}
	 \bm{W} = \bm{A}+\frac{\alpha}{n}\bm{11^T},
	 \label{eq:transition_prob}
\end{equation}
where we set $\alpha=0.15$ and $\bm{1}$ is a vector of all ones from $\mathbb{R}^{n}$.
Finally, we normalize each row in $\bm{W}$ to sum up to $1$ to obtain the transition matrix $\bm{P}$.

Now, the stationary distribution $\bm{\pi}$ can be calculated as the left eigenvector of $\bm{P}$ with its largest eigenvalue $1$ and the stationary distribution satisfies the eigenvalue equation for the matrix $\bm{P}$: $\bm{\pi^T}=\bm{\pi^T P}$.
An example calculation of the stationary distribution (without the teleportation factor) is depicted in Table~\ref{tab:sd_example}. Assuming two users exhibit the same number of page views on the pages A, B and C, we can incorporate the information about the chronological ordering of the sequence by calculating the stationary distribution. Hence, a sequence of ``ABCABC'' page visits (top row of Table~\ref{tab:sd_example}) yields a different stationary distribution than a sequence of ``AABBCC'' page visits (bottom row of Table~\ref{tab:sd_example}).

\begin{figure*}[!ht]
\centering
\subfigure{\includegraphics[width=1\textwidth]{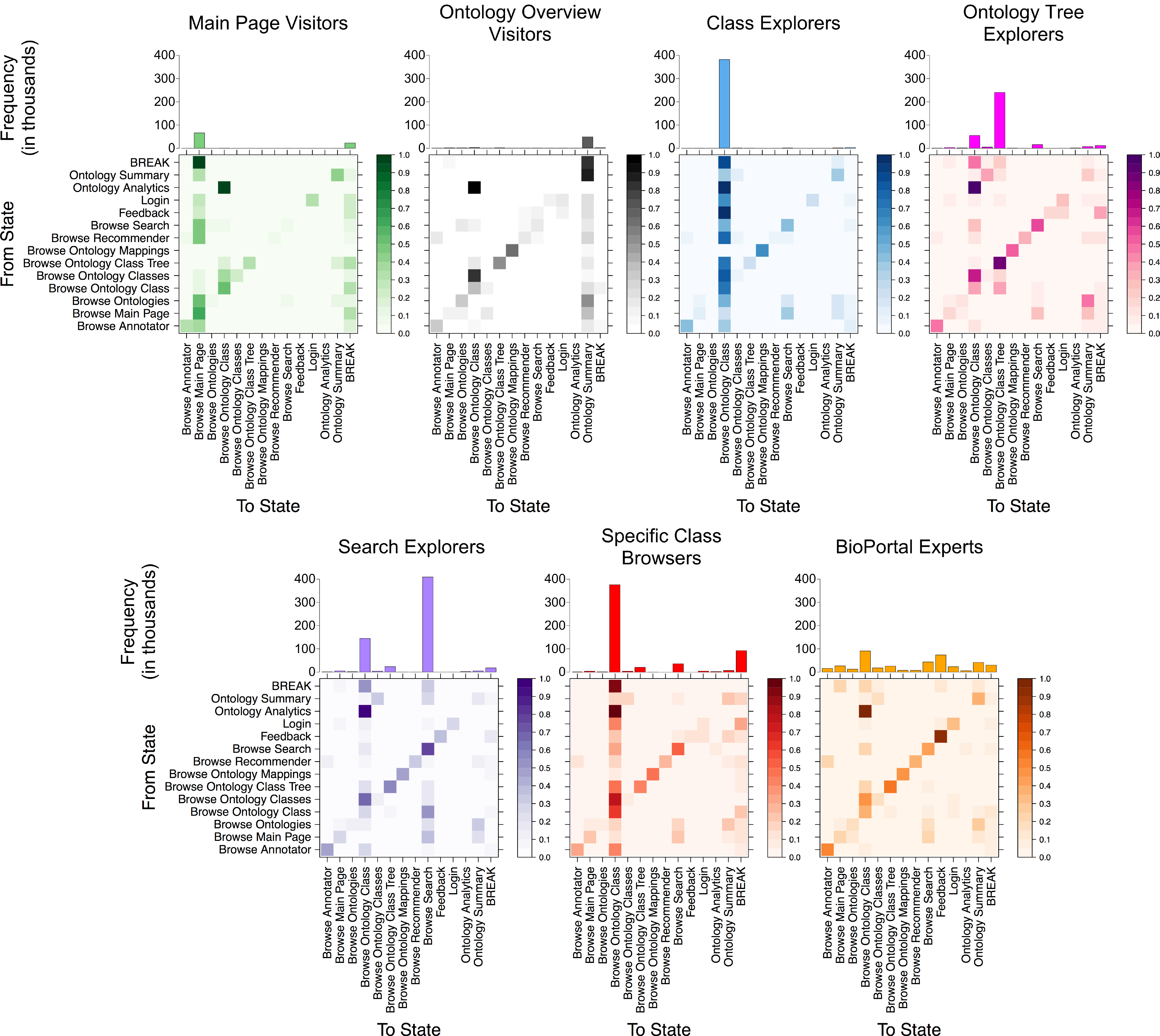}}
\caption{
\textbf{\textit{Browsing-Behavior Types}:} We have extracted and visualized the transition probabilities between the different actions for all $7$ clusters obtained by $K$-means when using the stationary distribution vectors (bottom). Transitions are always read \textit{From State} (left) to \textit{To State} (bottom) and rows are normalized individually. The darker the colors, the higher the transition probability between two states. Histograms of the actions are depicted on top of the transition matrices and indicate their absolute occurrences in the extracted sequences for each cluster. All of the clusters exhibit differences in the frequencies (histograms) and transition probabilities.%
}
\label{fig:bp_uclusers}
\vspace{-5pt}
\end{figure*}

\subsection{Clustering Browsing Behavior}
\label{subsec:clustering}

To deepen our understanding about the browsing behavior of users on BioPortal, we first group similar users. To that end, we use $K$-means---an unsupervised clustering al\-go\-rithm---which clusters ``close-by'' users. For $K$-means we need to embed users in a vector space so that their proximity can be determined by, for example, pairwise Euclidean distances between user vectors. For that purpose we represent users with their stationary distribution vectors. 

Once, when users are represented as vectors, $K$-means randomly selects $K$ initial central vectors (centroids), where $K$ has to be manually set and defines the number of clusters the algorithm will find. 
In every iteration, each user vector is assigned to the closest centroid. The positions of the centroids are then updated (i.e., moved towards the center of the corresponding vectors). This process continues until convergence, that is until the displacement of the centroids between iterations is below a certain threshold.
Note that the choice of the clustering algorithm is arbitrary and we leave it open to future work, to determine which algorithm is best suited for a specific application. For illustration purposes, we have chosen $K$-means as it is well studied and can handle large datasets ($168,008$ distinct users, clustered over $34$ action-label dimensions).

\section{Results}

\subsection{Differences in User Browsing Behavior}
\label{subsec:res:clusters}

We have calculated the stationary distribution for each user as described in Section~\ref{subsec:clustering}, and used $K$-means to group them accordingly. Additionally, we clustered users using static page view vectors.%

\noindent \textbf{Estimating \& Validating $K$.} To estimate a plausible number of clusters, which best represent the different types of browsing behavior in BioPortal, we have calculated the percentage of variance explained by $1$ to $25$ clusters for our empirical data (cf. Figure~\ref{fig:bp_kval}). This method is also often referred to as ``elbow''-method, and states that the optimal value for $K$, given empirical data, should be set so that the introduction of additional clusters only minimally increases the explained variance. For both our experiments, the clustering via the stationary distributions and page views, the best value for $K$ is $7$.

To be able to visually inspect and validate the resulting clusters, we have extracted three principle components by applying PCA on the stationary distribution vectors and visualized the results (see Figure~\ref{fig:bp_statdist_clusters}). Each point represents one user of our dataset and the colors indicate one of the seven clusters suggested by $K$-means. Additionally, we have added (i) the action labels with the largest and smallest coefficients (see Table~\ref{tab:pca coeffs}) to the corresponding axes to allow for manual interpretation and (ii) provide labels for the different clusters. 
We can see that the clusters obtained by $K$-means, when using the stationary distribution vectors, are easily distinguishable and warrant further inspection.

\begin{table}[!b]
\vspace{-15pt}
\center
\footnotesize
\caption{\textit{Principal Component Coefficients.} The explained variance for the first principal component (PC1) is $63.67\%$, for the second (PC2) $73.57\%$ and $80.53\%$ for the third (PC3).}
\begin{tabular}{ l | c | c | c }\toprule
\multicolumn{1}{c|}{\textbf{Action Label}} & \textbf{PC1}& \textbf{PC2}& \textbf{PC3}  \\\midrule
Browse Search & $-0.1897$ & $+0.9088$ & $-0.0368$ \\
Ontology Summary & $-0.1253$ & $-0.2736$ & $+0.6240$ \\
Browse Main Page & $-0.1236$ & $-0.2785$ & $-0.7712$ \\
Browse Ontology Class & $+0.9571$ & $-0.0702$ & $-0.0127$ \\\bottomrule
\end{tabular}
\label{tab:pca coeffs}
\end{table}

\begin{figure*}[!t]
\centering
\subfigure[Ontology Projects over Principal Components]{\label{fig:bp_top50_pca}\includegraphics[width=0.5\textwidth]{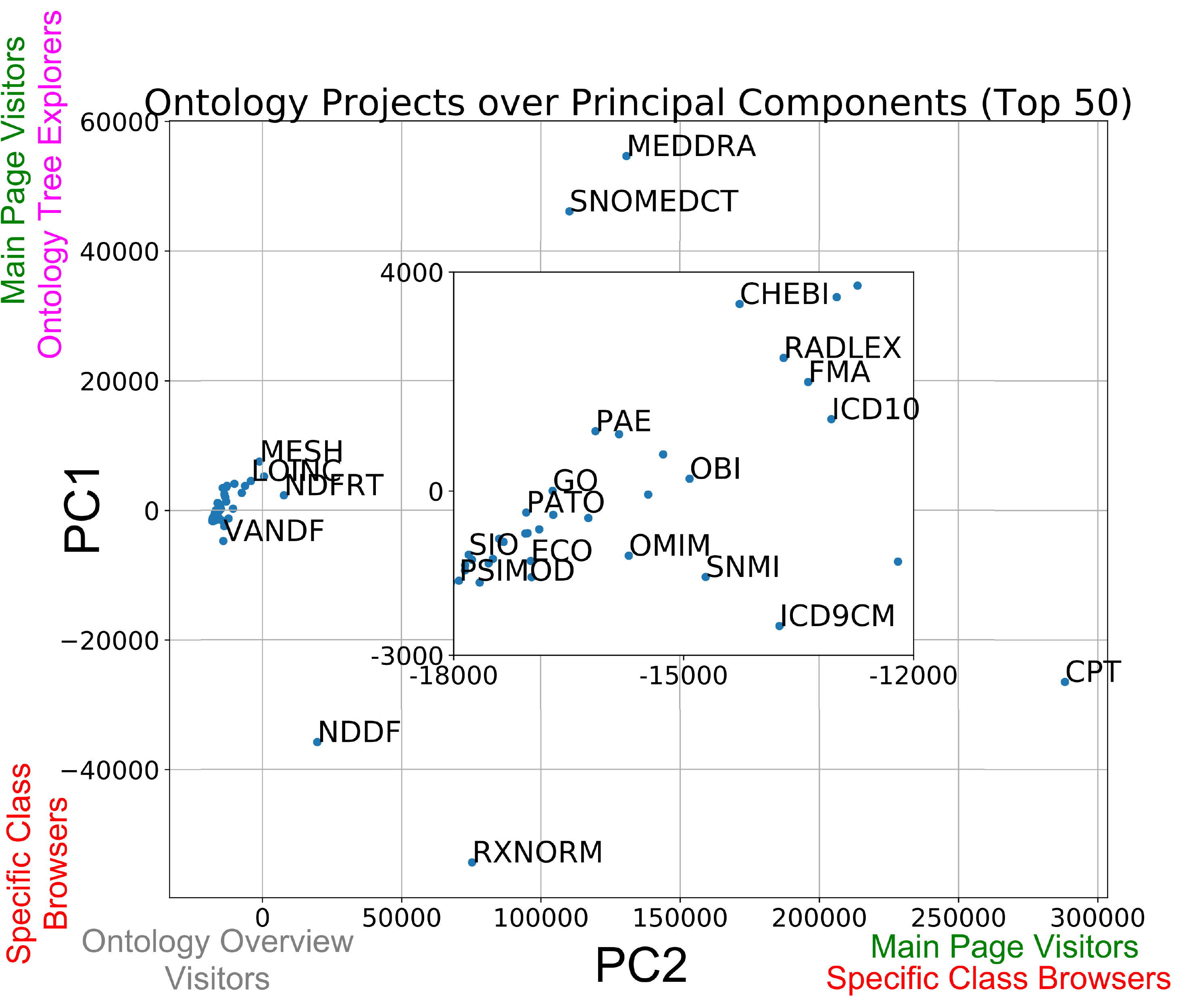}}
\subfigure[Comparison of Exploration Behavior (CPT \& RXNORM)]{\label{fig:bp_browsing_comp:rel}\includegraphics[width=0.4\textwidth]{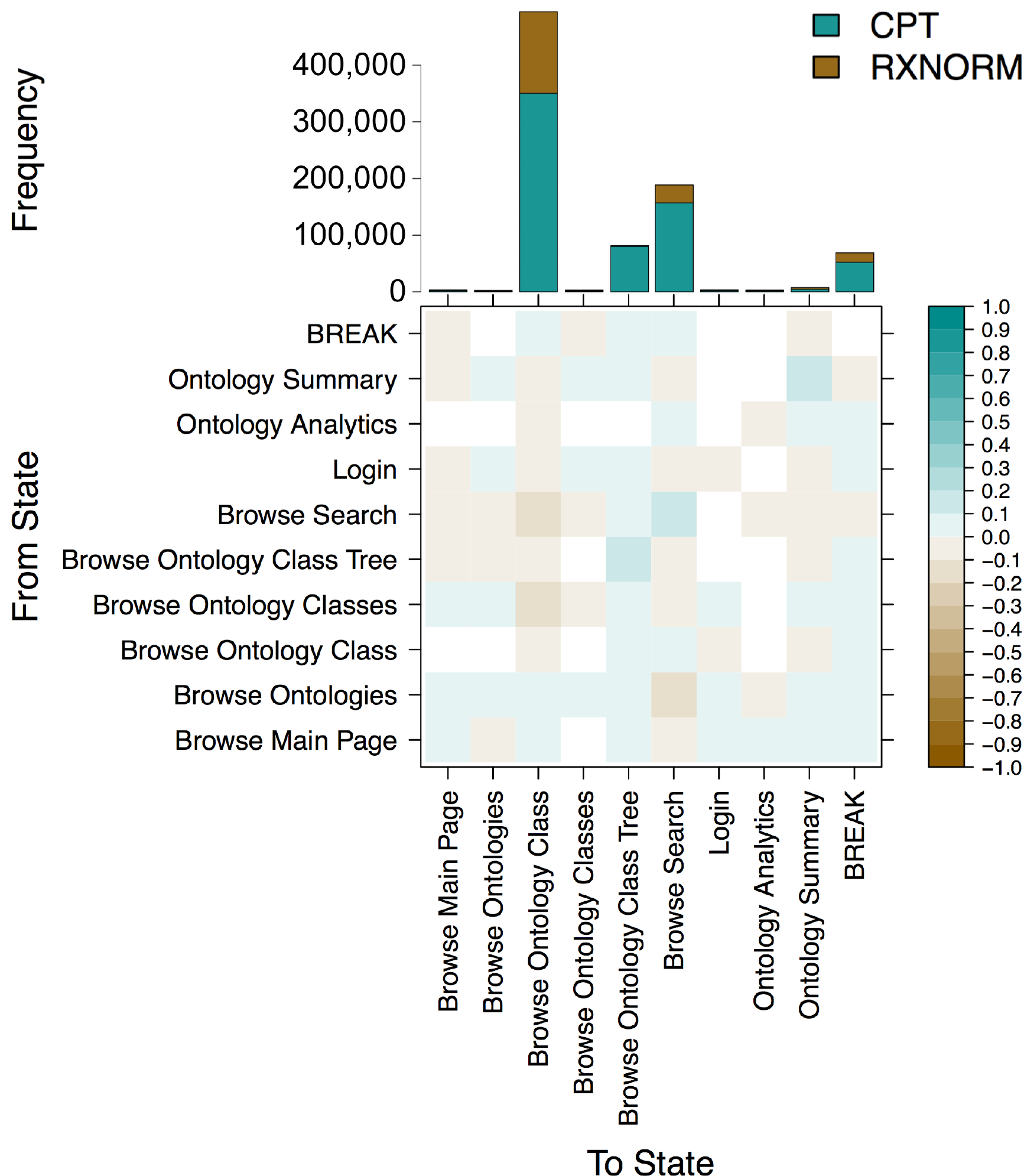}}
\caption{
\textbf{\textit{Comparison of Exploration Behavior}.} We have aggregated the number of actions per browsing-behavior cluster, applied PCA and plotted the landscape of exploration behavior types for the $50$ most visited ontologies (blue points) in (a). The extremes of the axes correspond to the clusters with the largest and smallest coefficients of PCA respectively. We have added labels for the ontologies and terminologies where appropriate. The inset magnifies the densely concentrated group of ontologies. The larger the difference in actions performed by different browsing-behavior clusters for each ontology, the larger the distance between ontologies. 
The detailed comparison of CPT (green) and RXNORM (brown) is depicted in (b). The histogram (top) depicts how often the different actions occurred in the extracted sequences in both ontologies. The transition matrix (bottom) depicts the importance of the transitions between actions of users while browsing CPT (blue) and RXNORM (brown). Transitions that are equally important in both projects are white. The transition from \textit{Browse Search} (From State) to \textit{Browse Ontology Class} (To State) is more dominant for RXNORM. Additionally, RXNORM exhibits a very small number of \textit{Browse Ontology Class Tree} actions.}
\label{fig:bp_browsing_comp}
\vspace{-7pt}
\end{figure*}

\noindent \textbf{Categorizing Clusters.} Hence, we have extracted and visualized the action sequences of all clusters (see Figure~\ref{fig:bp_uclusers}; the colors correspond to the clusters in Figure~\ref{fig:bp_statdist_clusters}) individually. Note that we have removed actions with very few occurrences from the visualizations for reasons of readability, which is why the transition probabilities of the rows in Figure~\ref{fig:bp_uclusers} do not necessarily sum up to $1.0$. 

Using the stationary distribution as input for $K$-means, we were able to obtain a total of $7$ distinguishable browsing-behavior types on BioPortal: 

\noindent The \textbf{Main Page Visitors} primarily visit the main page of BioPortal and exhibit very few other actions. The cluster consists of $2,813$ users with an average of $32.2$ actions (median of $2$) per user. 

\noindent  The \textbf{Ontology Overview Visitors} consist of $2,668$ users, who primarily visit ontology overview pages and refrain from exploring the classes of the ontologies. The average user in this group performs $22.7$ actions (median of $3$).

\noindent The group of \textbf{Class Explorers}, the biggest cluster with a total of $109,159$ users, specifically target and visit classes of the different ontologies hosted on BioPortal. The average number of actions for this group is $3.6$ (a median of $2$). 

\noindent \textbf{Ontology Tree Explorers} use the hierarchical representation of an ontology on BioPortal to explore and browse the content of an ontology. A total of $4,490$ users belong to this group with an average of $77.3$ actions (median of $14$).

\noindent A total of $6,701$ \textbf{Search Explorers} make extensive use of the search functionality provided by BioPortal to identify, explore and find classes in the ontologies. On average, users of this cluster conduct $92.1$ actions (median of $18$).

\noindent \textbf{Specific Class Browsers} represent the second largest brow\-sing behavior type with $22,436$ users. In contrast to \textit{Search Explorers}, this group exhibits shorter browsing sessions, evident in the higher number of \textit{BREAK} states, and concentrates on exploring multiple classes of a specific area of an ontology (as opposed to whole ontologies, by inspecting the ontology class tree). The average number of actions for users in this group is $24.9$ with a median of $8$.

\noindent The final group are \textbf{BioPortal Experts}, who use several features that BioPortal provides to find and explore ontologies and classes, such as the annotator and recommender. On average, the $19,741$ users of this cluster performed $24$ actions (median of $4$).

\noindent\textbf{Comparison to Page Views.} We repeated the same experiment with the static page view vectors as input for $K$-means and manually inspected the results for $K=7$ clusters. In contrast to the results shown in Figure~\ref{fig:bp_statdist_clusters}, we were only able to distinguish a total of four different kinds of browsing-behavior types. The biggest group of users are the \textit{Search Explorers}, consisting of a total of $167,979$ users and a total of four different clusters with $167,631$, $1$, $21$ and $326$ users respectively. We consider them as single cluster as the actions and transition matrices of these groups were nearly indistinguishable upon manual inspection. The three remaining clusters, \textit{Class Explorers}, \textit{Ontology Tree Explorers} and \textit{Feedback Providers}, are again very small, with only $3$, $6$ and $20$ users respectively.

\noindent\textbf{Increasing $K$.} Additionally, we have manually explored up to $14$ clusters obtained from $K$-means, using the stationary distributions as input. Due to the information inherent in the stationary distribution, we could further refine the obtained class labels. In additional to the seven clusters outlined before, we identified \textit{Annotators} ($542$ users), \textit{Widget Browsers} ($399$ users), \textit{Search \& Tree Explorers} ($3,685$ users), \textit{Feedback Providers} ($9,311$ users) and \textit{Account Maintainers} ($374$ users).

\subsection{Differences when Exploring Ontologies}
\label{subsec:res:browsing_comparison}

To be able to learn more about exploration and usage strategies of users, we were interested in identifying differences in the distribution of browsing-behavior types that interact with the different ontologies on BioPortal. 
Hence, we have determined the number of users for each browsing-behavior type cluster for the $50$ most visited ontologies in our dataset and aggregated the corresponding number of actions. Finally, we have applied PCA on the aggregated numbers of requests per cluster and visualized the results (similar to Figure~\ref{fig:bp_statdist_clusters}) to identify differences in the browsing-behavior types between projects (see Figure~\ref{fig:bp_top50_pca}). 
The larger the difference in actions performed by different browsing-behavior clusters for each project, the larger the distance between the projects.

Due to limitations in space, we restrict the presentation of results to two projects.
In particular, we compare the browsing behavior of users for the \textit{Current Procedural Terminology} (CPT) as well as \textit{RxNorm} (RXNORM). The former is a terminology, which is maintained by the American Medical Association, and is used to describe services in the medical, surgical, and diagnostic domain (mostly exclusive) for billing purposes in the US. The latter was developed by the National Library of Medicine as a public resource for various applications, and represents a terminology that contains all medications available on the US market. 
We have selected CPT and RXNORM, as they are among the ontologies that have received the most visits, exhibit substantially different rankings for the actions per browsing-behavior cluster and were initially designed for different practical applications.

\noindent \textbf{Sequence Extraction.} For this analysis, we have extracted all sequences of all users, who performed at least $20$\% of all their actions on one of the corresponding projects. As a consequence, sequences are not mutually exclusive, meaning that one user could be present in both of our extracted action sequences if that user performed at least $20\%$ of all clicks on CPT and another $20\%$ on RXNORM. Given that the majority of users only visit one or two ontologies (see Figure~\ref{fig:bp_timediff:ont_per_user}) and that there are many actions, which can not be assigned to an ontology, $20\%$ of all actions already represents a very high filtering criterium.
Overall, we collected a total of $61,367$ unique IP addresses for CPT, which conducted on average $10.7$ actions (median of $2$). In contrast, we extracted action sequences of $23,757$ IP addresses for RXNORM, with an average of $8.6$ actions (median of $2$). The aggregated actions for each cluster for the two ontologies are listed in  Table~\ref{tab:cluster counts}.

\noindent \textbf{Comparing Browsing-Behavior Types.} To visualize and investigate the differences in browsing-behavior strategies between the two ontologies, we have fit a first-order Markov chain on the extracted sequences of both projects and visualized the difference between the two transition matrices in Figure~\ref{fig:bp_browsing_comp:rel}. Note that we only display the top $10$ most frequent actions for reasons of readability. 

CPT---being the more popular ontology---received roughly three times more requests than RXNORM, as shown in the histogram on top of Figure~\ref{fig:bp_browsing_comp:rel}. The most common actions for both projects are \textit{Browse Ontology Class} and \textit{Browse Search}. In contrast to RXNORM, users of CPT also rely on the hierarchy of the ontology to browse classes (see \textit{Browse Ontology Class Tree} in Figure~\ref{fig:bp_browsing_comp:rel}). This is not surprising, as RXNORM is a flat ontology, which does not have a hierarchy that can be visualized by BioPortal. This also explains why the relative importance of the transition from \textit{Browse Search} to \textit{Browse Ontology Class} is more dominant for RXNORM, than it is for CPT (evident in the dark brown transition probabilities in Figure~\ref{fig:bp_browsing_comp:rel}).

\begin{table}[!b]
\vspace{-17pt}
\center
\tiny
\caption{\textit{Actions for CPT \& RXNORM.}}
\begin{tabular}{ l  r  r }\toprule
\multicolumn{1}{c}{\textbf{User Type}} & \textbf{CPT} Actions & \textbf{RXNORM} Actions \\\midrule
Class Explorers & $120,926$ ($3$) & $49,440$ ($2$)\\
Specific Class Browsers & $187,987$ ($2$) & $88,068$ ($1$)\\
BioPortal Experts & $19,934$ ($5$) & $12,755$ ($4$)\\
Main Page Visitors & $195,030$ ($1$) & $35,869$ ($3$)\\
Ontology Tree Explorers & $78,485$ ($4$) & $1,478$ ($5$)\\
Ontology Overview Visitors & $252$ ($7$) & $69$ ($7$)\\
Search Explorers & $1,142$ ($6$) & $224$ ($6$)\\\bottomrule
\end{tabular}
\label{tab:cluster counts}
\end{table}

\section{Discussion}

In this paper we have analyzed, modeled and clustered the browsing behavior of users on BioPortal. In particular, we have grouped users according to their browsing behavior in Section~\ref{subsec:res:clusters} and compared how users explore ontologies in Section~\ref{subsec:res:browsing_comparison}. 

\noindent\textbf{Differences in Browsing-Behavior Types.} We have shown that we were able to obtain multiple, clearly distinguishable browsing-behavior types, when using stationary distribution as input for $K$-means. Further, we have calculated the explained variance per cluster to determine the number of browsing-behavior types to investigate. The manual inspection of the obtained clusters shows that users primarily concentrate on browsing specific classes (\textit{Class Explorers} and \textit{Specific Class Browsers}) of biomedical ontologies on BioPortal. Two strategies of users to explore ontologies involve the exploitation of the ontological hierarchy (\textit{Ontology Tree Explorers}; if available) and the search functionality provided by BioPortal (\textit{Search Explorers}). Users in the third-largest cluster ($19,741$)---BioPortal Experts---use advanced functionality, such as the annotator or the recommender, to find classes and explore ontologies. 

The comparison of our results with clusters obtained from static page views indicate that the inclusion of information about the dynamic nature of a user's interaction behavior with a website is important. %

In general, our results indicate that users exhibit a preference to explore and browse the content of the classes of biomedical ontologies. However, this also indicates that additional efforts are warranted to promote and further enhance and evaluate the utility of the functionality provided by BioPortal to explore and find ontologies. For example, only a small portion of users (the BioPortal Experts) use the mappings between ontologies. 

\noindent\textbf{Differences when Exploring Ontologies.} Our results indicate that several characteristics of ontologies, such as the hierarchical structure (or lack thereof), can influence how users explore and interact with these semantic vocabularies and terminologies on BioPortal. One particular example of such an ontology is RXNORM, which does not exhibit a structured hierarchy, and is thus not displayed in the class explorer on BioPortal. The only ways to interact with classes of this ontology on the website is to either use the search functionality, which requires prior knowledge about the content of the ontology, or to rely on mappings from other ontologies. However, this also explains the strong focus of users/visitors of RXNORM to use the search functionality (\textit{Browse Search}) to explore, investigate and find classes of this ontology on BioPortal (see Figure~\ref{fig:bp_browsing_comp:rel}). 

In contrast, the structural hierarchy of CPT is used by BioPortal to visualize the class tree. Hence, to explore the ontology, users can (and do)---aside from searching (\textit{Browse Search}) for specific classes---use and explore the hierarchy (\textit{Browse Ontology Class Tree}) to find a specific class. Note that we have only presented the differences in browsing-behavior types for two of more than $500$ ontologies! Yet, we have shown that simple differences (i.e., the lack of a class hierarchy) between projects cause users of BioPortal to employ different strategies to find classes or explore the ontological content.

\section{Conclusions \& Future Work} 

In this paper, we have presented an approach for identifying different types of user behavior in ontology exploration log data using stationary distributions from first order Markov chains. We provide evidence for a total of $7$ distinct browsing-behavior types on BioPortal and offer an interpretation of their relevance and meaning. In addition, we cluster ontologies by their user behavior and identify pragmatic commonalities among ontology projects. Our results advance our understanding of the ways in which ontological repositories, such as BioPortal, are used and may guide the development of more user-oriented systems for ontology exploration and reuse on the Web. 

For future work, we are particularly interested in further elaborating the comparison between the browsing behavior of different ontologies on BioPortal. In doing that, we might be able to identify commonalities for groups of ontologies that trigger differences in the way users interact with ontologies on BioPortal. Understanding the different ways in which semantic structures may influence ontology exploration behavior has the potential to help in the design of BioPortal as well as in the development of more effective and more reusable ontologies.

\section{Acknowledgments}
This work was supported by Grant GM086587 from the U.S. National Institute of General Medical Sciences (NIGMS) of the National Institutes of Health. The Prot\'eg\'e project is supported by NIGMS grant GM103316. NCBO is funded by National Institutes of Health grant U54 HG004028.

\bibliographystyle{abbrv}

\end{document}